# On k-hypertournament losing scores

Shariefuddin Pirzada
Department of Mathematics
University of Kashmir, India
King Fahd University of Petroleum
and Minerals, Saudi Arabia
email: sdpirzada@yahoo.co.in

Guofei Zhou
Department of Mathematics
Nanjing University, Nanjing, China
email: gfzhou@nju.edu.cn

**Abstract.** We give a new and short proof of a theorem on k-hypertournament losing scores due to Zhou et al. [8].

## 1  Introduction

An edge of a graph is a pair of vertices and an edge of a hypergraph is a subset of the vertex set, consisting of at least two vertices. An edge in a hypergraph consisting of k vertices is called a k-edge, and a hypergraph all of whose edges are k-edges is called a k-hypergraph.

A k-hypertournament is a complete k-hypergraph with each k-edge endowed with an orientation, that is, a linear arrangement of the vertices contained in the hyperedge. In other words, given two non-negative integers $n$ and $k$ with $n \geq k > 1$, a k-hypertournament on $n$ vertices is a pair $(V, A)$, where $V$ is a set of vertices with $|V| = n$ and $A$ is a set of k-tuples of vertices, called arcs, such that any k-subset $S$ of $V$, $A$ contains exactly one of the k! k-tuples whose entries belong to $S$. If $n < k$, $A = \phi$ and this type of hypertournament is called a null-hypertournament. Clearly, a 2-hypertournament is simply a tournament. Let $e = (v_1, v_2, \ldots, v_k)$ be an arc in a k-hypertournament H. Then $e(v_i, v_j)$ represents the arc obtained from $e$ by interchanging $v_i$ and $v_j$.

The following result due to Landau [5] characterises the score sequences in tournaments.

---

**Theorem 1** *A sequence of non-negative integers $[s_1, s_2, \ldots, s_n]$ in non-decreasing order is a score sequence of some tournament if and only if for $1 \leq j \leq n$*

$$\sum_{i=1}^{j} s_i \geq \binom{j}{2},$$

*with equality when $j = n$.*

Now, there exist several proofs of Landau's theorem and a survey of these can be found in Reid [6]. Brualdi and Shen [1] obtained inequalities on the scores in tournaments which are individually stronger than that of Landau, but collectively the two are equivalent. Although tournament theory has attracted many graph theorists and much work has been reported in various journals, the latest can be seen in Iványi [2].

Instead of scores of vertices in a tournament, Zhou et al. [8] considered scores and losing scores of vertices in a k-hypertournament, and derived a result analogous to Landau's theorem [5]. The score $s(v_i)$ or $s_i$ of a vertex $v_i$ is the number of arcs containing $v_i$ and in which $v_i$ is not the last element, and the losing score $r(v_i)$ or $r_i$ of a vertex $v_i$ is the number of arcs containing $v_i$ and in which $v_i$ is the last element. The score sequence (losing score sequence) is formed by listing the scores (losing scores) in non-decreasing order.

For two integers $p$ and $q$,

$$\binom{p}{q} = \frac{p!}{q!(p-q)!}$$

if $p \geq q$ and

$$\binom{p}{q} = 0$$

if $p < q$.

The following characterisation of losing score sequences in k-hypertournaments is due to Zhou et al. [8].

**Theorem 2** *Given two non-negative integers $n$ and $k$ with $n \geq k > 1$, a non-decreasing sequence $R = [r_1, r_2, \ldots, r_n]$ of non-negative integers is a losing score sequence of some k-hypertournament if and only if for each $j$,*

$$\sum_{i=1}^{j} r_i \geq \binom{j}{k}, \qquad (1)$$

*with equality when $j = n$.*

## 2 New proof

Koh and Ree [4] have given a different proof of Theorem 2. Some more results on scores of k-hypertournaments can be found in [3, 7]. The following is the new and short proof of Theorem 2.

**Proof.** The necessity part is obvious.

We prove sufficiency by contradiction. Assume all sequences of non-negative integers in non-decreasing order of length fewer than $n$, satisfying conditions (1) are losing score sequences. Let $n$ be the smallest length and $r_1$ be the smallest possible with that choice of $n$ such that $R = [r_1, r_2, \ldots, r_n]$ is not a losing score sequence.

Consider two cases, (a) equality in (1) holds for some $j < n$, and (b) each inequality in (1) is strict for all $j < n$.

**Case (a).** Assume $j$ ($j < n$) is the smallest such that

$$\sum_{i=1}^{j} r_i = \binom{j}{k}.$$

By the minimality of $n$, the sequence $[r_1, r_2, \ldots, r_j]$ is the losing score sequence of some k-hypertournament $H_1$. Also

$$\sum_{i=1}^{m} [r_{j+i} - \frac{1}{m} \sum_{i=1}^{k-1} \binom{j}{i}\binom{n-j}{k-i}] = \sum_{i=1}^{m+j} r_i - \binom{j}{k} - \sum_{i=1}^{k-1}\binom{j}{i}\binom{n-j}{k-i}$$

$$\geq \binom{m+j}{k} - \binom{j}{k} - \sum_{i=1}^{k-1}\binom{j}{i}\binom{n-j}{k-i}$$

$$= \binom{m}{k},$$

for each $m$, $1 \leq m \leq n-j$, with equality when $m = n-j$.

Let

$$\frac{1}{m} \sum_{i=1}^{k-1} \binom{j}{i}\binom{n-j}{k-i} = \alpha.$$

Therefore, by the minimality of $n$, the sequence

$$[r_{k+1} - \alpha, r_{k+2} - \alpha, \ldots, r_n - \alpha]$$

is the losing score sequence of some k-hypertournament $H_2$. Taking disjoint union of $H_1$ and $H_2$, and adding all $m\alpha$ arcs between $H_1$ and $H_2$ such that each arc among $m\alpha$ has the last entry in $H_2$ and each vertex of $H_2$ gets equal shares from these $m\alpha$ last entries, we obtain a k-hypertournament with losing score sequence R, which is a contradiction.

**Case (b).** Let each inequality in (1) is strict when $j < n$, and in particular $r_1 > 0$. Then the sequence $[r_1 - 1, r_2, \ldots, r_n + 1]$ satisfies (1), and therefore by minimality of $r_1$, is the losing score sequence of some k-hypertournament H, a contradiction. Let x and y be the vertices respectively with losing scores $r_n + 1$ and $r_1 - 1$. If there is an arc e containing both x and y with y as the last element in e, let $e' = (x, y)$. Clearly, $(H-e) \cup e'$ is the k-hypertournament with losing score sequence R, again a contradiction. If not, since $r(x) > r(y)$ there exist two arcs of the form

$$e_1 = (w_1, w_2, \ldots, w_{l-1}, u, w_l, \ldots, w_{k-1})$$

and

$$e_2 = (w'_1, w'_2, \ldots, w'_{k-1}, v),$$

where $(w'_1, w'_2, \ldots, w'_{k-1})$ is a permutation of $(w_1, w_2, \ldots, w_{k-1})$, $x \notin \{w_1, w_2, \ldots, w_{k-1}\}$ and $y \notin \{w_1, w_2, \ldots, w_{k-1}\}$. Then, clearly R is the losing score sequence of the k-hypertournament $(H - (e_1 \cup e_2)) \cup (e'_1 \cup e'_2)$, where $e'_1 = (u, w_{k-1})$, $e'_2 = (w'_t, v)$ and t is the integer with $w'_t = w_{k-1}$. This again contradicts the hypothesis. Hence, the result follows. □